\documentclass[12pt,preprint,longnamesfirst,epsf]{aastex}

\newcommand {\kms}{km~s$^{-1}$}
\newcommand {\msol}{$M_{\scriptstyle\odot}$}

\begin{document}

\title{X-ray Emission from Wind Blown Bubbles.  III. ASCA SIS Observations
of NGC~6888}

\author{Matthias Wrigge\altaffilmark{1}, You-Hua Chu\altaffilmark{2},
Eugene A. Magnier\altaffilmark{3}, Heinrich J. Wendker\altaffilmark{1}}
\altaffiltext{1}{Hamburger Sternwarte, Gojenbergsweg 112, D-21029 Hamburg,
 Germany; mwrigge@t-online.de, hjwendker@hs.uni-hamburg.de}
\altaffiltext{2}{Astronomy Department, University of Illinois, 1002
 W. Green Street, Urbana, IL 61801;  chu@astro.uiuc.edu}
\altaffiltext{3}{Canada-France-Hawaii Telescope Corporation, P.O. Box 1597, 
  Kamuela, HI 96743; magnier@cfht.hawaii.edu}

\begin{abstract} 
We present {\it ASCA} SIS observations of the wind-blown bubble NGC~6888.  
Owing to the higher sensitivity of the SIS for higher energy photons
compared to the {\it ROSAT} PSPC, we are able to detect a $T \sim 
8\times10^6$ K plasma component in addition to the $T \sim 1.3\times10^6$ K 
component previously detected in PSPC observations.
No significant temperature variations are detected within NGC\,6888.
Garc\'{\i}a-Segura \& Mac Low's (1995) analytical models of WR bubbles 
constrained by the observed size, expansion velocity, and mass of the 
nebular shell under-predict the stellar wind luminosity, and cannot
reproduce simultaneously the observed X-ray luminosity, spectrum, surface 
brightness profile, and SIS count rate of NGC\,6888's bubble interior.
The agreement between observations and expectations from models can be 
improved if one or more of the following ad hoc assumptions are made:
(1) the stellar wind luminosity was weaker in the past, (2) the
bubble is at a special evolutionary stage and the nebular shell has 
recently been decelerated to 1/2 of its previous expansion velocity, 
and (3) the heat conduction between the hot interior and the cool 
nebular shell is suppressed.
{\it Chandra} and {\it XMM-Newton} observations with high spatial 
resolution and high sensitivity are needed to determine accurately
the physical conditions NGC~6888's interior hot gas for critical 
comparisons with bubble models.
\end{abstract}

\clearpage

\section{Introduction}

The fast stellar wind from a massive star can sweep up its ambient 
medium into a dense shell up to a few $\times$10 pc across, with
the central cavity filled with a hot, tenuous, X-ray-emitting 
plasma, a so-called wind-blown bubble \citep[e.g.,][]{Wea77}.
It is expected that every star with sufficient 
wind power should be surrounded by a wind-blown bubble.

Wolf-Rayet (WR) stars are excellent candidates for producing bubbles, 
as they have winds with high terminal velocities, $>$ 1000 \kms,
and large mass-loss rates, 10$^{-5}$--10$^{-4}$ \msol~yr$^{-1}$
\citep{PBH90}.  Optical surveys 
of WR ring nebulae \citep{CTK83,HBG82,MC93,MCG94,Metal94}
show that about 1/4 of the $\sim$150 observed Galactic WR stars are
associated with ring-like nebulae.  Most of these nebulae, however,
are amorphous rings with dynamic ages much larger than the lifetime
of a WR phase, and hence they have been suggested to be merely 
photoionized, instead of being dynamically shaped, by the WR stars
\citep{Chu81}.  Only about 10 WR ring nebulae have sharp 
rims and short dynamical ages suggesting that they are bubbles blown 
by the central stars during the WR phase.  The small number of WR bubbles 
indicates that their formation mechanism may not be as simple as one 
naively thinks.

The morphology and dynamical structure of a wind-blown bubble are 
dependent on the distribution of the ambient medium.  
\citet{Wea77} derived analytical solutions 
to the dynamical evolution of a bubble in a homogeneous medium, and 
showed that heat conduction and evaporation across the contact 
discontinuity which separates the hot interior from the outer shell 
determines the temperature and density profiles of the interior.  
Their work was later complemented by the numerical calculations of 
\citet{Roz85}.
These early calculations were later extended by \citet[][hereafter
GM]{GM95a} analytically and by \citet{GM95b} numerically to a more 
realistic model incorporating the time dependence of the wind parameters
due to the evolution of the central star.  
According to the evolutionary sequence of 
massive stars \citep[][and references therein]{CM86,vdH92}, the
progenitor of a WR star is a luminous blue variable 
(LBV; M $\gtrsim$ 60~\msol) or a red 
supergiant (RSG; 25~\msol\ $\lesssim$ M $\lesssim$ 60~\msol).  In the 
latter case, the expansion of the fast WR wind into the slow RSG 
wind with a $r^{-2}$ density profile naturally leads to dynamic 
instabilities (Vishniac instability) in the swept-up shell.
As the swept-up shell advances past the outer edge of the RSG wind, 
further instabilities (Rayleigh-Taylor instability) set in
and fragment the shell.  Thus the dense shell becomes
clumpy and filamentary, allowing the hot shocked stellar wind to 
break out and form a shock ahead of the fragmented dense shell
\citep[][hereafter GLM]{GLM96}.
This model has been successful in describing the 
basic optical morphology of bubbles blown by WR stars.

Of the $\sim$10 WR stars with wind-blown bubbles, four have been
observed in soft X-rays with the {\it ROSAT} Position Sensitive Proportional 
Counter (PSPC): NGC~2359, NGC~3199, NGC~6888, and S\,308.  
The latter two have been detected, but their observed X-ray properties 
are discrepant from the theoretical predictions in two respects (Wrigge, 
Wendker, \& Wisotzki 1994, hereafter Paper I; Wrigge 1999, hereafter 
Paper II). 
First, the observed X-ray luminosity is more than a factor of 10
lower than that expected in models constrainted by independently 
observed stellar wind velocity, mass-loss rate, and bubble dynamics,
such as shell size, expansion velocity, mass, etc.\ 
\citep[e.g.,][]{BL85}.
Second, the X-ray surface brightness profiles (in terms of the 
observed count rate distribution) disagree with predictions.  Both 
NGC~6888  and S 308 show limb-brightened surface brightness 
profiles as opposed to the center-filled appearance expected in
models.

Paper I suggested two possible explanations for the X-ray surface
brightness profiles:  
(1) The ring nebula might be in an advanced
evolutionary stage when the dense swept-up shell has already
overtaken the entire RSG wind, leading to the shell
fragmentation and allowing the hot gas in the bubble interior 
to expand faster into the surrounding medium.
Thus, the inner stellar wind shock front could advance closer 
to the outer edge of the shell and reduce the volume of hot gas;
furthermore, the temperature and density of the hot gas could be
lowered due to the fast adiabatic expansion.
(2) The WR shell has not actually overtaken the entire RSG wind, 
but the heat conduction across the contact discontinuity 
is less effective than assumed in the analytical models
because of the saturation of heat fluxes \citep{CM77,DB93} or 
the presence of magnetic fields \citep[e.g.,][]{Spi67}. 
In this case, a bubble interior would have higher temperatures 
and lower densities, which make the X-ray emission difficult to 
detect because of the relatively soft energy bandpass of the 
{\it ROSAT} PSPC.

To distinguish between these two possibilities, X-ray observations 
with a higher spectral resolution and a harder energy bandpass
are needed.  We have therefore obtained observations of 
NGC~6888 using the {\it Advanced Satellite for Cosmology and 
Astrophysics} ({\it ASCA}).  This paper reports our analysis of 
these observations.  In Section 2 we describe 
the {\it ASCA} SIS observations and data reduction. 
Section~\ref{results+c} describes the spectral fitting results 
and derives physical properties of the hot gas inside the 
bubble. The observations are compared with model predictions in 
Section 4, and a summary is given in Section 5.

\section{X-ray Observations}

We have carried out X-ray observations of NGC~6888 with the {\it ASCA
X-ray Observatory}.  
Our goal was to use the improved spectral resolution and extended 
spectral coverage of {\it ASCA} to further our understanding of the
physical conditions of the hot gas interior to NGC~6888, in light of 
the discrepancies between results of {\it ROSAT} PSPC observations 
and model predictions reported in Paper I.
We have used both the {\it ROSAT} PSPC and {\it ASCA} observations in 
the analysis reported in this paper. 

The {\it ROSAT} PSPC observations have been discussed in detail in Paper I.
We present here only a brief summary. The observations were 
performed on 1991 April 13 -- 14, and are archived under the 
observation number WP900025.  For these observations, the WR star
WR\,136 (HD\,192163) was centered in the PSPC.  The on-axis angular
resolution of the PSPC was $\sim$ 30\arcsec\ at 1 keV.  The PSPC 
is sensitive in the energy range of 0.1 -- 2.4 keV and has an 
energy resolution of $\sim$ 43\% at 1 keV.  The total effective 
exposure time was 7,866~s.  The PSPC image of NGC\,6888, smoothed 
to 80$''$ resolution, is shown in Figure 1a.

Our {\it ASCA} observations were performed on 1995 October 31 -- November 1. 
The {\it ASCA Observatory} has been described by \citet{TIH94}. 
Briefly, the satellite carries four imaging thin-foil grazing incidence
X-ray telescopes.  Two of the telescopes are focused on Solid-state
Imaging Spectrometers, called SIS~0 and SIS~1.  The other two telescopes
use Gas Imaging Spectrometers (GIS) as the detectors.  The GIS
observations of NGC\,6888 are not used in our analysis because of the
GIS's low sensitivity to photon energies below $\sim$ 1 keV, where
most X-ray emission from NGC~6888 concentrates.  The SIS detectors 
have an energy resolution of $\sim$ 2\% at 6 keV and cover the energy 
range 0.4 -- 10 keV, with reduced throughputs near the ends of the band.
The SIS's point spread function has a narrow core of $\sim$ 1\arcmin\ 
diameter and a half-power radius of 3\arcmin.  Each of the two SIS 
detectors is composed of 4 separate CCDs.  Not all CCDs must be activated 
for a given observation.  The presence of ``hot'' and ``flickering'' 
pixels, which fill the telemetry with false signals, has made it 
necessary to use only a subset of each detector for many observations. 

Because of the telemetry limitations, our observations were performed
in the two-CCD mode.  We used complementary pairs of CCDs from the 
SIS~0 and SIS~1 detectors so that the entire field of view of the 
SIS was covered, although any point on the sky was observed by only
one of the two telescopes.  The center of the SIS detector mosaic was 
placed at approximately the geometric center of the optical shell of 
NGC~6888.  The observations were 
performed in the faint mode when allowed by the telemetry, and 
converted to the bright mode on the ground for analysis.  Using the
standard processing software ({\em ascascreen} and {\em ftools}), we
removed bad time periods using the following selection criteria: aspect
deviation $<$ 0.01$^\circ$, angle to bright earth $> 20^\circ$, 
elevation $>$ 10$^\circ$, minimum cutoff rigidity of 6, and PIXL rejection
of 75.  We also removed hot and flickering pixels.  After screening, a
total of 43,308~s of usable data remained for SIS~0 and 39,936~s
remained for SIS~1.  We included only high and medium bit-rate data 
because of problems reported for low bit-rate data.  A smoothed
SIS image of NGC~6888 is shown in Figure 1b.

We defined two sets of source regions to study the X-ray emission
from NGC~6888.  The first set consists of a pair of rectangular 
regions centered on the two bright emission peaks, similar to 
those defined in Paper I; these two regions will be called the North 
and South Regions.  The second set comprises three elliptical 
regions concentric with the optical shell of NGC~6888.  These three
elliptical regions allow us to study the variation of X-ray 
properties with radius.  These regions are outlined in Figure 1b.

An important issue is the determination of the background spectra.
For the PSPC, we extracted a background spectrum from an annulus at
the outer edge of the inner portion of the detector.  This annulus is
free of emission from NGC~6888 and contains only one obvious source,
which was excluded in the extraction (see Fig.~1).  It is necessary 
to correct this background spectrum for the energy-dependent 
vignetting in the instrument.  For the PSPC, the vignetting 
is a function of radius and energy.  We have determined the ratio of the
effective areas as a function of energy for each of the regions.  This
ratio is nearly constant for energy $\lesssim$ 1.5 keV, where the
majority of PSPC background photons originate.  Because the ratio is
nearly constant, we have applied only a constant correction, which is
estimated to introduce no more than 2\% error on the background flux.
Since the background flux is always a factor of 4 to 5 lower than the
source flux, this does not introduce a significant error into the PSPC
spectra.

For the {\it ASCA} SIS data, there are two ways to determine the 
background.  It is possible to use a source-free region in the SIS 
image as the background, similar to what we have done with the PSPC 
data; however, there are several problems with this approach.  
First, the energy dependence of the effective area correction is 
larger.  Second, the smaller field of view of the SIS means that 
the source-free background regions have a smaller total amount of 
signal and hence will introduce noises in the background subtraction.
Finally, in our SIS observations of NGC~6888, a large source-free 
region is found in one active CCD of the SIS~1 detector, but not 
the other active CCD of SIS~1 or either of the active CCDs of SIS~0.
Since the detectors have different channel-to-energy 
relationships, it is not advised to apply background from one 
detector to another.  The alternative to this background subtraction 
is to use data from the background field observations supplied by 
the {\it ASCA} Guest Observer Facility.  These background observations
are taken in source-free regions in the sky, and are sorted according
to the cutoff rigidity of the observations.  We have selected a field 
appropriate for our observations, and double-checked the validity of 
this background field for our region by subtracting it from a 
source-free region in our field.  Only 16\% of the flux remains, and 
roughly half of it is in the highest 50 energy channels that are not 
included in our spectral fits.  The error introduced by this chosen
background field is probably small, but may not be negligible for the
1--2 keV band that is crucial in the spectral fits.
Background spectra for the source regions are extracted from the 
background field using the same region templates.

For the {\it ASCA} data, spectra were extracted from both SIS~0 and SIS~1.
The datasets from the two SIS telescopes and the PSPC cannot be simply
combined directly (i.e., added together) as the three telescopes have
distinct response matrices.  Instead, the three spectra were fitted
jointly to the same models using XSPEC.  The normalizations were
fitted as independent free parameters to allow for differences in
spatial and spectral resolution and the different portions of the sky
covered by the regions.  Spectral bins for all three instruments were
grouped together in order to give sufficient counts in each spectral
bin so that the $\chi^2$ statistics would be valid.  We have used
primarily spectral models of optically thin thermal plasmas from 
\citet[][hereafter RS]{RS77}.  In some cases, as discussed 
below, Mewe-Kaastra models \citep[][hereafter MEKA]{MGV85,KA92}
were also used for comparisons with RS models.

\section{\label{results+c}Results} 

\subsection{The X-ray Image}

Figure 1 shows X-ray images of NGC~6888 in the 0.4--2.4 keV band from 
(a) the {\it ROSAT} PSPC and (b) the {\it ASCA} SIS 
observations.  To improve the signal to noise ratio, both the PSPC 
and SIS images were smoothed to an effective angular resolution of 
80\arcsec\ (FWHM).  The smoothed images show very similar X-ray 
morphologies with the X-ray emission distributed in a shell-like 
structure just interior to the optical filaments.  No diffuse 
emission from the central region of the bubble is detected.  For 
the SIS observations discussed below, the detection threshold is 
$8.8\times10^{-4}$ counts~s$^{-1}$~arcmin$^{-2}$, corresponding to 
a surface brightness 3$\sigma$ above the background of 
$4.8\times10^{-4}$ counts~s$^{-1}$~arcmin$^{-2}$.

\subsection{X-ray Spectra}
\subsubsection{North and South Regions}

In order to compare the {\it ASCA} observation with the previous {\it 
ROSAT} observations, we extracted SIS spectra from the same North and
South Regions used in Paper I for the PSPC observations.  Figure 2 shows 
the combined count rate spectra from the PSPC (thin crosses) and SIS 
(thick crosses) datasets for (a) the North and (b) the South Regions.  
The SIS detected 191 and 360 net source photons in these two regions, 
respectively.  The SIS count rate for the two regions combined is 
1.4$\times10^{-2}$ counts s$^{-1}$.   Since thermal plasma emission 
is the most probable emission mechanism in wind blown bubbles,
we fitted RS models of optically thin ionization-equilibrium plasma
emission to the data.  The free parameters in the fits are the plasma
temperature $T$, the absorption column density parameterized by the 
hydrogen column density $N_{\rm H}$, and the normalization factor 
$A = 10^{-14} (\int n_{\rm e}^2 dV) / (4\pi D^2)$, where $n_{\rm e}$ is
the electron density of the plasma, $V$ is the volume, and $D$ is the
distance.  In some of the fits, $T$ or $N_{\rm H}$ may be held fixed.

The small number of counts detected does not warrant elaborate
spectral models; therefore, we start with a single temperature
plasma model.
We find that the combined PSPC + SIS count rate spectra cannot be
satisfactorily fitted with a single temperature plasma model.  
An additional high temperature component contributing fluxes to the 
high energy channels is needed to yield acceptable reduced 
$\chi^2$ values.  
This marked difference from the PSPC observations (Paper I) is 
mainly due to the higher energy resolution of the SIS detector.
We have also varied the abundances of the elements N, Ne, Fe, and Ni, 
which contribute to the line emission in the SIS band, but the fits 
do not generally improve significantly, unless we increase the abundances 
to unphysically high values.  We therefore have fixed the abundances to 
solar values.  We have also tried MEKA models for the fits.  Generally
the derived plasma parameters are compatible with those found with RS
fits within the error limits, but it is worthwhile noting that the derived
fluxes and luminosities differ by up to 30\%.  This may be considered
a lower limit of the uncertainties associated with the values.

One important goal of this work is to identify any variations in the
observed plasma temperature as a function of position, corresponding
to variations in the physical conditions of the emitting plasma.  To
investigate temperature variations across the nebula, it is crucial that 
the foreground absorption column density is accurately determined.  In 
the soft X-ray regime, one often finds the fitted temperatures in the 
$(1 - 5)\times10^6$~K range to be strongly dependent on the fitted 
$N_{\rm H}$ values: lower temperatures can be compensated by higher 
absorption column densities (see the analysis of S\,308 in
Paper II for an example). 
For this reason, both the absolute value of $N_{\rm H}$ and its possible 
variation across the nebula must be known. For the latter, \citet{Wen75} 
compared H$\alpha$ and 21-cm radio continuum observations of NGC~6888 and 
showed that $A_{\rm V}$ is constant over the nebula.  To compare the optical 
and X-ray extinctions, we fitted two-temperature plasma emission models 
to the spectra of the North and South Regions, and calculated the $\chi^2$ 
contour plots of $N_{\rm H}$ versus the lower temperature (the higher 
temperature and normalization as the free parameters), shown in Figure 3.  
The acceptable values of $N_{\rm H}$ are in the range 
$(3 - 3.8)\times10^{21}$ H cm$^{-2}$.  The reddening towards WR 136, 
the central star of NGC~6888, was found to be $E(b-v) = 0.45$ mag from 
optical measurements \citep{HWK94}.  Using the Galactic gas-to-dust ratio 
$N_{\rm H}$/$E(B-V)$ = $5.8\times10^{21}$ H cm$^{-2}$ mag$^{-1}$
\citep{BSD78} and $E(B-V) = 1.21 E(b-v)$ \citep{LS80},
the reddening can be converted to 
$N_{\rm H}$ = $3.13\times10^{21}$ H cm$^{-2}$.  This value was also 
adopted in Paper I.  The optical and X-ray measurements of $N_{\rm H}$ 
agree well.  We thus use a fixed value of $N_{\rm H}$ = 
$3.13\times10^{21}$ H cm$^{-2}$, with an uncertainty of 30\%, for 
the spectral analysis.  
The fitted spectra are plotted in Figure~2 as solid lines, and the 
results of the fits are listed in Table~1.

One immediate result is that the temperatures of the low- and
high-temperature components do not change significantly from the
North to the South Region (see Table~1). 
The ratio of the normalization factors, $A$, for the lower and higher 
temperature components increases by a factor of 1.8 from the 
North to the South Region. Since the cooling
function (in the SIS band) does not change significantly over the
temperature range of interest, the relative contribution of the lower
temperature plasma to the total flux from these regions increases by
the same factor.  Note, however, that the errors in the fit parameters
are large, so these results are also compatible with the hypothesis
that the North and South Regions have similar thermal properties.
This has also been deduced from the {\it ROSAT} PSPC data.  Owing to the 
sensitivity of the SIS to higher energy photons, we are now able to 
extend this conclusion to gas in the temperature range 
$(1 - 10)\times10^6$~K.

Table~1 also lists the observed and
absorption-corrected X-ray fluxes.  The observed PSPC + SIS
fluxes agree very well with the results in Paper I, but our absorption
corrected fluxes are higher by a factor of 4. This is mainly due to
the slightly lower temperature we found for the low-temperature
component: for a given foreground absorption column density, the
factor converting observed count rate to unabsorbed flux increases 
with decreasing temperature.

\subsubsection{Elliptical Regions}

The analysis of the North and South Regions suggests that there 
is no discernible temperature differences between the two ends along 
the major axis of NGC~6888.  We next search for radial temperature
variations by analyzing spectra extracted from concentric elliptical 
annuli. 
The elliptical regions are shown in Figure~1b.   We define
three regions, Ellipses 1, 2, and 3, corresponding to the outermost 
annulus, the middle annulus, and the innermost ellipse, respectively.
For each region, a PSPC spectrum is extracted from the entire region,
and two SIS spectra are extracted separately from the two portions
covered by the SIS~0 and SIS~1.  The three spectra (PSPC, SIS~0, and
SIS~1) are fitted simultaneously; to account for the different volumes
actually sampled by each spectrum, three normalization factors for 
each temperature component are fitted as free parameters.

We have performed spectral fits incorporating single-temperature 
RS models as well as two-temperature RS models.  Ellipses 1 and 
2 require two temperatures in order to fit the data well, but
Ellipse 3 shows only a marginal improvement in the $\chi^2$ with two
temperatures.  It should be noted, however, that Ellipse 3 has a low
number of counts, which may explain this difference.  As shown in
\S3.2.1 for the rectangular regions, which encompass most of the 
same fluxes, it is not possible to fit a single-temperature 
RS spectrum by allowing the abundances to vary, even with significant 
changes in certain elements.  Thus, for the fits in the elliptical
regions, we have set the abundances to the solar value, and fixed 
the absorption column density to $3.13\times10^{21}$ H cm$^{-2}$, the 
value implied by the optical extinction.  Table~2 shows the results 
of fits for the three elliptical regions.  Instead of listing all 
18 normalization factors, we give only the sum of the normalization 
factors for the SIS spectra.  Note that within the errors the sum of 
the SIS normalization factors agree with the PSPC normalization 
factor, confirming the consistency of the simultaneous fits.

Again, we do not find any significant temperature variation among the
elliptical regions for either the low- or high-temperature component.
Owing to the large uncertainty in the normalization factors, it is not
possible to estimate variations in the relative contributions of the
low- and high-temperature components to the X-ray emission from 
these regions.

\section{Discussion}

Previous X-ray studies of NGC~6888 compared observations with 
Weaver et al.'s models and found large discrepancies
\citep[Paper I;][]{Boc88}.  It is now understood that Weaver et 
al.'s model of interstellar bubbles in a homogeneous medium is
not applicable, because NGC~6888 was formed by the interaction 
between the WR wind and a previous RSG wind with a radial density
drop-off proportional to $r^{-2}$.
NGC~6888 is better described by GM's models 
taking into account the RSG wind.  Thus, we compare the  {\it ASCA} 
observations with predictions from the analytical models of GM.

We have searched the literature for independent observations of 
physical parameters of the stellar wind and nebular shell of NGC~6888.
The best determined physical parameters as well as the relevant 
references are summarized in Table~3.  
Using GM's analytical models, a bubble can be fully specified 
if the size, mass, and expansion velocity of the nebular shell
are known.
We thus use the observed parameters of the nebular shell and
GM's model to produce the density and temperature profiles
in the bubble interior, and further use a thermal emission model as 
implemented in the \citet{RS77} code (kindly provided by J.~Raymond)
to calculate the expected X-ray luminosity, spectra, and surface 
brightness profiles of NGC~6888 for comparison with observations.
In addition, we use the nebular shell parameters and the GM model 
to determine the expected mechanical luminosity of the stellar wind
($L_{\rm w}$), which can also be compared to that derived from 
the observed stellar wind terminal velocity ($v_\infty$), and mass
loss rate ($\dot M$), i.e., $L_{\rm w} = (1/2) \dot M v_\infty^2$,
as another independent check of the model.

We have considered six GM models with different shell masses and
expansion velocities (see Table 4) for the following reasons.
The ionized shell mass of NGC\,6888 has been estimated to be 
4 $M_\odot$ \citep{Wen75} and the neutral shell mass 
$\sim$40 $M_\odot$ \citep{MM88}.  We consider these two masses
as the lower and upper limits of the shell mass of NGC\,6888;
thus models 1, 3, and 5 use 4 $M_\odot$ as the shell mass, 
and models 2, 4, and 6 use 40 $M_\odot$ as the shell mass.
To explore the parameter space, we have used the observed expansion
velocity of 75 km~s$^{-1}$ in models 1 and 2, and assumed expansion
velocities of 100 km~s$^{-1}$ in models 3 and 4, and 200 km~s$^{-1}$
in models 5 and 6.
The expected stellar wind luminosity, X-ray luminosity of the 
bubble interior, and {\it ASCA} SIS count rate are listed in Table 4.

The observed {\it ASCA} SIS spectra and surface brightness profiles 
are affected by interstellar absorption, the energy- and 
position-dependent effective area of the X-ray telescope (XRT), 
and the energy-dependent detector response matrix (DRM).  It is 
not possible to remove these effects from the observed spectra 
and surface brightness profiles to derive intrinsic properties
for comparison with model predictions.
It is necessary to apply these absorption and instrument 
effects to the model predictions and compare the simulated 
{\it ASCA} SIS spectra and surface brightness profiles to the 
observations.

For the interstellar absorption cross section we used the 
analytical representation of the photoelectric cross section as 
given by \citet{MM83}.  The effective area of the {\it ASCA} XRT 
we used is tabulated in the file ``xrt\_ea\_v2\_0.fits" provided 
in the {\it ASCA} calibration database at {\it legacy.gsfc.nasa.gov}. 
Note that the effective area file incorporates an azimuthal angle
dependence for the effective area.  Since the deviation from the 
mean effective area is only $\lesssim$ 5\% in the energy range 
relevant for our observations, we have omitted the azimuthal 
dependence and used the effective area for $\phi = 0^\circ$.  
Concerning the DRM, we note that the {\it ASCA} SIS actually consists 
of two separate detectors, SIS~0 and SIS~1 with four CCDs each 
with different DRMs.  Since the differences in the DRMs are 
small compared to the uncertainty in the emission models, we used 
the DRM for SIS~0 chip~1, file ``s0c1g0234p40e1\_512v0\_8i.rmf" 
from the {\it legacy} database, in the simulations.
The simulated {\it ASCA} SIS surface-brightness profiles and spectra 
for the six models are shown in Figures~4 and 5, respectively.
In Figure~4, lines are drawn to mark the surface brightness that
would be detected with 10$\sigma$ for an aperture of 1 arcmin$^2$
in our {\it ASCA} SIS observations.  (This 10$\sigma$ threshold 
is selected to represent an ``easy detection".)

Compared to the observed $L_{\rm w}$ = $5\times10^{37}$ ergs 
s$^{-1}$, $L_{\rm x} \sim 9\times10^{34}$ ergs s$^{-1}$, and
{\it ASCA} SIS count rate = $6.6\times10^{-2}$ counts s$^{-1}$
(summed over the three elliptical regions), model 1 fails
completely, under-predicting these luminosities and count rate 
by 1--2 orders of magnitude.
Artificially raising the shell expansion velocity to 100 km~s$^{-1}$,
model 3 still misses the marks by more than an order of magnitude.
Using a high nebular shell mass, 40 $M_\odot$, models 2 and 4 
produce reasonable $L_{\rm x}$, but over-predict the {\it ASCA} SIS 
count rate and under-predict $L_{\rm w}$ by an order of magnitude.

The incompatibility of models 1--4 with observations is also evident
in spectral shape and surface brightness profile.
The observed SIS spectra show two emission maxima whereas the expected
spectra from models 1 and 3 exhibit only the lower energy peak, 
indicating  a lack of a very hot plasma component in models.
Indeed the best model fits to these two simulated spectra 
yield temperatures of only $\sim1.5\times10^6~{\rm K}$ whereas 
the observed spectrum indicates an additional hot component at 
$T$ $\sim$ $8\times10^6$~K. 
Models 2 and 4 produce reasonable spectral shapes, but their
surface brightness profiles are markedly different from that
observed in NGC\,6888.  
As shown in Figure~1, NGC\,6888 has a limb-brightened morphology 
with hardly any emission above our detection limit in the central 
region.
However, models 2 and 4 predict a centrally filled appearance 
of the bubble resulting from the energy dependence of the foreground 
absorption.
While the emissivity per unit volume increases outwards (due to
the increasing density), the X-ray emission from the outer parts is
softer (due to the decreasing temperature) and more heavily absorbed
(due to the larger absorption cross section) especially for 
temperatures below $1.5\times10^6$~K.

%

We may explore whether it is possible to reproduce the observed X-ray 
properties of NGC~6888 by adjusting, within their uncertainties, the 
input parameters of the model.  As explained above, the foreground 
absorption column density $N_{\rm H}$ governs the observed surface 
brightness profile.  Since $N_{\rm H}$ is approximated by the \ion{H}{1} 
column density derived from the observed optical extinction and an 
empirically determined gas-to-dust ratio, a 30\% uncertainty is expected
\citep{BSD78}.  However, even if the adopted $N_{\rm H}$ = 
$3.13\times10^{21}$ H cm$^{-2}$ is increased or decreased by a factor of
3, the expected X-ray morphology of NGC~6888 does not change significantly.
The predicted count rates are roughly a factor of 3 higher and lower, 
respectively.


We next explore whether the uncertainties in observations of stellar 
wind properties may alleviate the discrepancy between bubble models 
and X-ray observations.
It has been suggested that the stellar wind mass loss rate may be
over-estimated by a factor of a few if the wind is clumpy \citep{MR94},
but the clumping-corrected mass loss rate of NGC\,6888's central star 
WR\,136 is very similar to the mass loss rate for a smooth wind,
$6.3\times10^{-5}$ $M_\odot$ yr$^{-1}$ \citep{NCW98}.
It is unlikely that the stellar wind luminosity was an order of magnitude
weaker in the past and gained strength only recently, as the mass loss rate
and stellar wind terminal velocity of WR 136 are not particularly high 
compared to other WR stars.
The observational properties of WR 136's stellar wind are probably
robust.

The hydrodynamical simulation of the formation and evolution of WR 
bubble by GLM offers an opportune instant that may solve the problem.
In their model specifically made for WR stars that have evolved through
a RSG phase, the WR shell becomes prominent in the optical when it 
collides with and overtakes the former RSG wind shell, and at the same 
time the WR shell decelerates to roughly 1/2 of its pre-collision 
velocity.  At this instant of time, the observed $v_{\rm exp}$ would 
be of the prominent but slower shell.
It is possible that NGC\,6888's WR shell has just overtaken the RSG 
shell so that the hot bubble interior is still in its adiabatic 
phase, and the temperature and density distributions have not changed
significantly from those of pre-collision when the shell expansion 
velocity was twice as high as the currently observed.
The numerical calculation from GLM for a 35 $M_\odot$ star,
similar to NGC~6888's central star WR~136, gives a pre-collision 
$v_{\rm exp}$ of 200 km~s$^{-1}$.
Thus, we have adopted a shell expansion velocity of 200 km~s$^{-1}$
in models 5 and 6.
Model 6, with a large nebular shell mass, produces an over-luminous
bubble, while model 5 produces reasonable $L_{\rm x}$ and surface 
brightness profile.
Model 5 appears to match observations the best among the six models
considered.

GLM's hydrodynamical simulation of WR bubble reproduces the observed 
bubble dynamics remarkably well, but does not yield a hot interior
with the observed X-ray surface brightness or luminosity.
The temperature of a bubble's hot interior depends not only on the
fast stellar wind's post-shock temperature but also on the heat 
conduction across the contact discontinuity between the hot interior 
and the cooler nebular shell.
It is possible that the heat conduction is not as efficient as 
assumed in models because of saturation of heat conduction 
\citep{DB93} and/or the presence of tangential magnetic field.
Therefore, we have kept the basic pressure-driven (or adiabatic)
bubble model, but made an ad hoc assumption that the heat conduction
rate is lower and approximated its effect by using a higher central 
temperature for the bubble interior.
The pressure of a bubble interior is given by the shell dynamics. 
We use Model 6 as the initial model since its expected $L_{\rm w}$
agrees best with the observationally derived $L_{\rm w}$.
We then artificially raise the central temperature by factors of 
2, 5, 10, and 20, and display the expected X-ray surface brightness
profiles in Figure~6.
The expected surface brightness profile for the second model,
with a central temperature of $4\times10^7$~K, shows a limb 
brightening (Fig.~6, upper-right panel), but its overall surface 
brightness is still considerably higher than that of the brightest 
peaks observed in NGC~6888.  Thus it seems that a simple increase 
of the temperature scale, or a decrease in the efficiency of heat 
conduction, goes in the right direction but drastic changes in the
conduction efficiency are needed to produce satisfactory surface 
brightness profiles.

Finally there may be a possibility that the X-ray emitting
volume is far smaller than predicted in the analytical model.
In fact, Paper I finds that the emitting volume in the {\it ROSAT} PSPC 
band comprises only $\approx$ 1\% of the total volume of the
bubble, but it was not clear whether there exists gas with 
temperatures of a few $\times10^7$~K.  This very hot gas is not detected
in the {\it ASCA} observations.
The numerical model of GLM shows that the outer RSG shell should
fragment and break apart shortly after the WR shell hit and overtook
it.  Qualitatively one would expect the hot interior of the bubble to
quickly expand through the gaps and cool adiabatically. In the meantime,
the inner shock front moves outward toward the shell and reduces 
the volume of hot gas. 
The {\it ASCA} SIS observations lack the angular resolution for detailed
analysis of the distribution of hot gas relative to the cool nebular
shell material.
Deep, high-resolution $Chandra$ or {\it XMM-Newton} observations 
will provide clear X-ray images for critical tests of bubble models.


\section{Summary and Conclusions}

We have obtained {\it ASCA} SIS observations of the wind-blown bubble
NGC~6888.  Its spectral coverage extends to higher energy range
than that of the {\it ROSAT} PSPC, thus allows us to detect a high
temperature component at $T \sim 8.5\times10^6$~K in addition to the
low temperature component at $T \sim 1.3\times10^6$~K.
Both components are detected near the periphery of the nebular shell.
No significant X-ray emission can be detected from the central regions
of NGC\,6888.  The X-ray emission has been divided into North and South
regions or three concentric elliptical annuli, but no significant 
temperature variations are detected.

GM's analytical models for bubbles produced by WR stars that are 
descendants of RSGs are used to simulate the observed X-ray spectrum,
surface brightness profile, and {\it ASCA} SIS count rate.
These observed parameters are compared with those expected from six 
GM models with two different nebular shell masses (4, 40 $M_\odot$)
and three different shell expansion velocities (75, 100, 200 km~s$^{-1}$).
None of these models produce all of the observed parameters satisfactorily.
Furthermore, none of these models can produce the limb-brightened 
X-ray morphology observed.  
To improve the agreement between observations and model expectations,
ad hoc assumptions are needed, such variable stellar wind luminosity,
recent deceleration of the nebular shell, and suppressed heat
conduction between the hot bubble interior and the cool nebular shell.

More observations and detections of wind-blown bubbles are needed
to gain a deeper understanding of the physics determining the soft 
X-ray emission from bubble interiors.  
It is disappointing and puzzling that diffuse X-ray emission has
been detected in only two WR bubbles \citep{Wri99}.  We have shown 
that besides 
integrated properties such as X-ray luminosity and observed count 
rate, the spatial (angular) variations of temperature, density and 
observed surface brightness are crucial for any comparison with
theoretical models. 
Deep X-ray observations with high angular and spectral resolution 
are needed for a detailed analysis of hot gas in NGC~6888.
Of particular importance is a clean excision of background point 
sources when extracting spectra of the diffuse X-ray emission.
Recent {\it XMM-Newton} observations of the WR bubble S~308 
have been used to demonstrate that the high-temperature plasma 
component derived from the {\it ROSAT} PSPC spectral fits 
(Paper II) may be an artifact owing to a contamination of 
unresolved point sources \citep{Cetal03}.
{\it Chandra} observations of the northern part of NGC~6888
have revealed numerous point sources \citep{Getal03}; a careful 
analysis is needed to verify or reject the high temperature 
component indicated by the {\it ASCA} SIS observations.
Sensitive {\it XMM-Newton} observations of NGC~6888 are needed
to verify the filamentary distribution of hot gas shown
in the {\it ROSAT} High Resolution Imager observations
\citep{WW02}.

\acknowledgements
We wish to thank J.\ Raymond for kindly providing unpublished
additions to the RS-model code.  We also thank R.\ Gruendl
and M. Guerrero for assistance in improving the figures.
MW was supported by DARA, grants FKZ 50 OR 9308 and 9604. 
YHC and EAM gratefully acknowledge the NASA grant NAG 5-2988.


\clearpage
\vskip 7cm

\begin{figure}   
\plotone{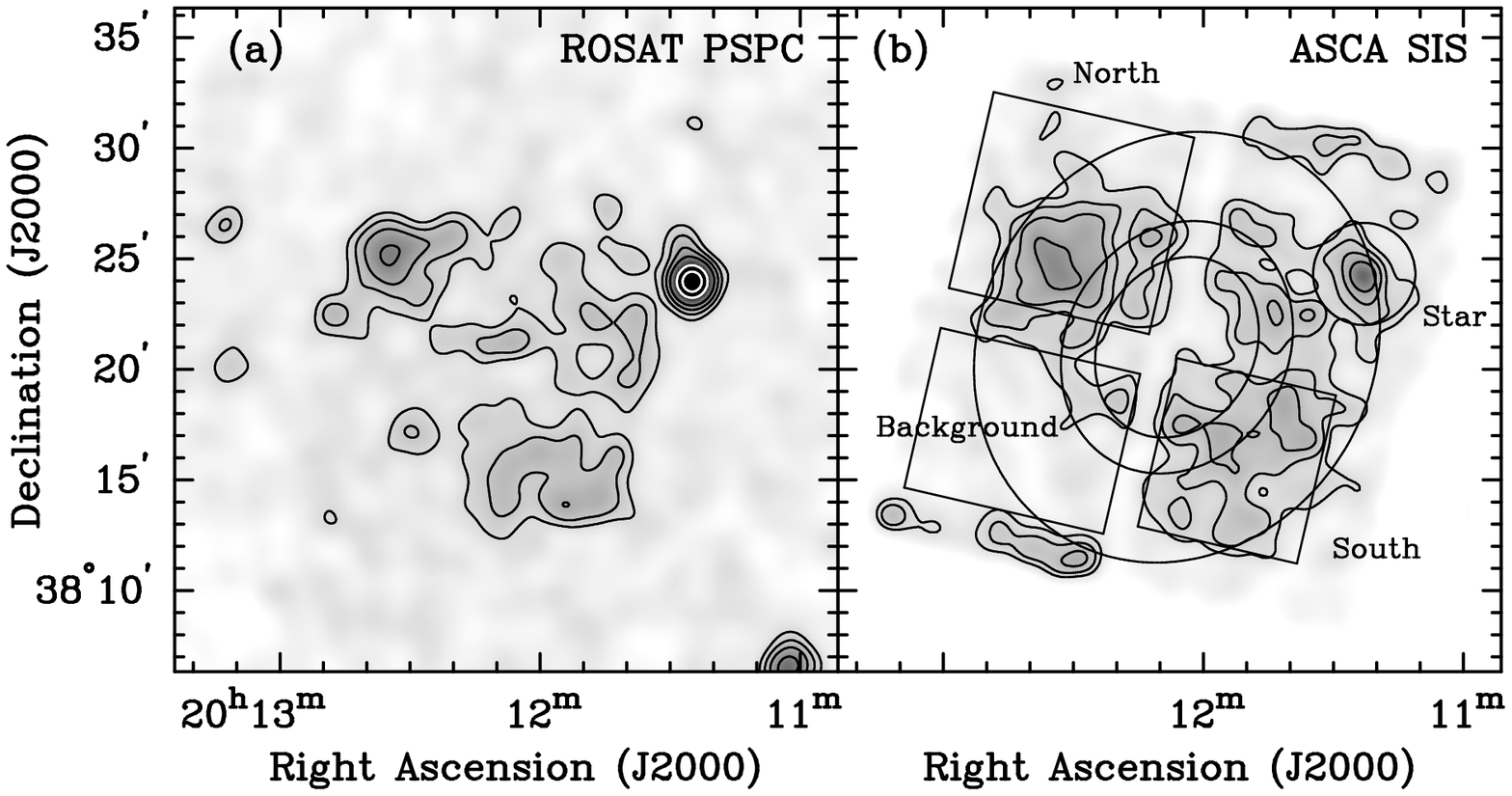}
\figcaption{\label{contur+f} (a) {\it ROSAT} PSPC image of NGC\,6888
in the 0.4--2.4 keV band, smoothed to an angular resolution of 80$''$.
Vignetting is not corrected.  The lowest contour level starts at
1.3$\times$10$^{-3}$ counts s$^{-1}$ arcmin$^{-2}$, 1.5$\sigma$ above 
the background of 8$\times$$10^{-4}$ counts s$^{-1}$ arcmin$^{-2}$.
The other contours are 3, 5, 8, 16, and 24 $\sigma$ above the 
background.  The two highest contours are plotted in white.
(b) {\it ASCA} SIS image of NGC~6888 in the same energy band and 
smoothed to the same angular resolution as that of the {\it ROSAT}
PSPC image. The first contour level corresponds to 6.8$\times$$10^{-4}$
counts s$^{-1}$ arcmin$^{-2}$, $1.5\sigma$ above the background of 
$\sim$4.8$\times$10$^{-4}$ counts s$^{-1}$ arcmin$^{-2}$. 
The other contours are 3, 5, and 8 $\sigma$ above the background.
Note that the apparent strip of low surface brightness corresponds
to the gap between the SIS\,0 and SIS\,1 detectors.
The bright point source to the northwest, marked by a circle, is a 
nearby G8V star HD\,192020 unrelated to the nebula.}
\end{figure}

\newpage

\begin{figure}  
\plottwo{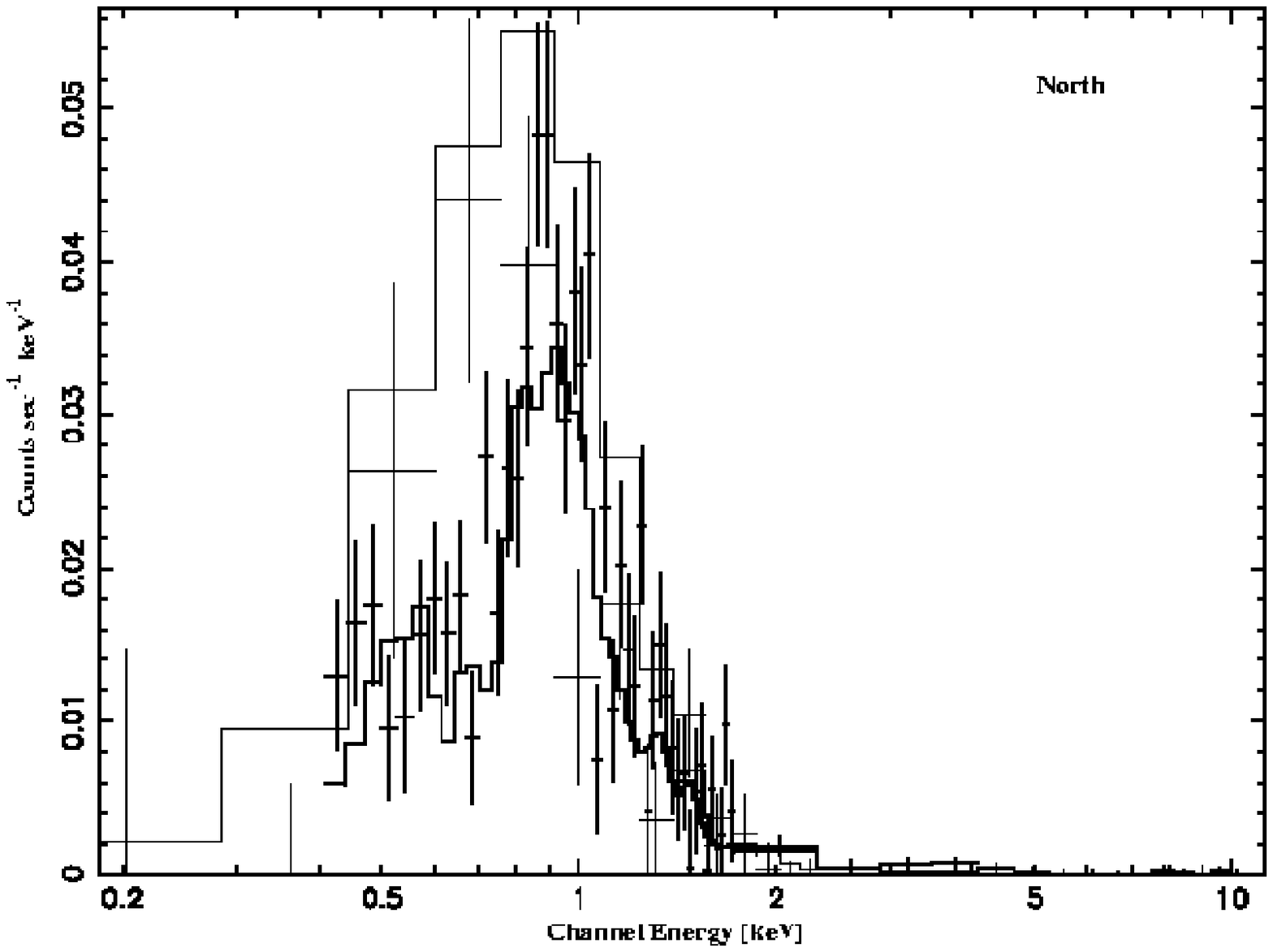}{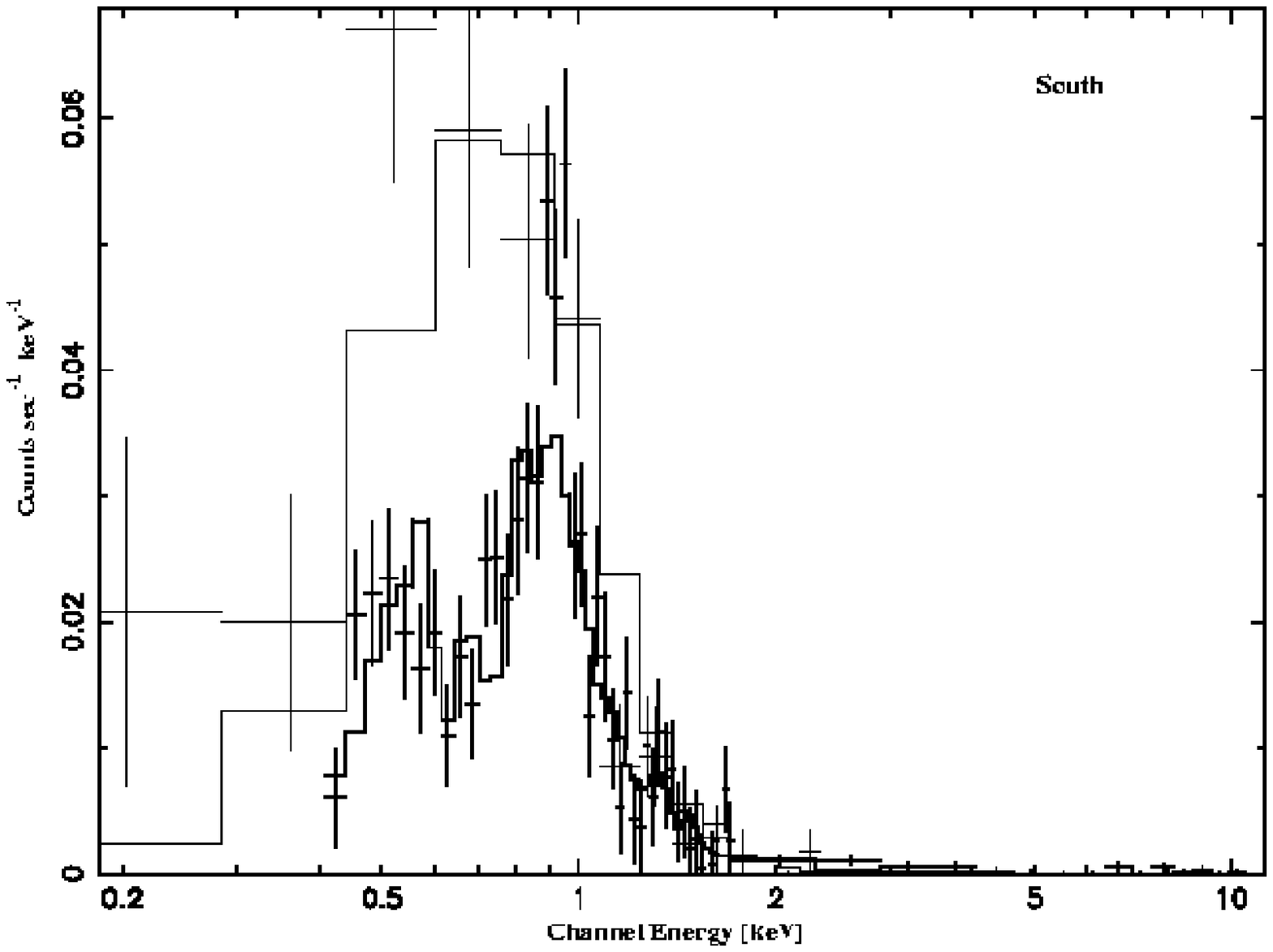}
\caption{\label{n-s-regions-fit+f} 
  Count rate spectra of (a) the North Region and (b) the South Region, 
  with the {\it ASCA} SIS data in thick crosses and
  the {\it ROSAT} PSPC data in thin crosses.  The PSPC and SIS
  spectra are fitted simultaneously and the best-fit is 
  plotted in solid lines.}
\end{figure}

\begin{figure}  
\plottwo{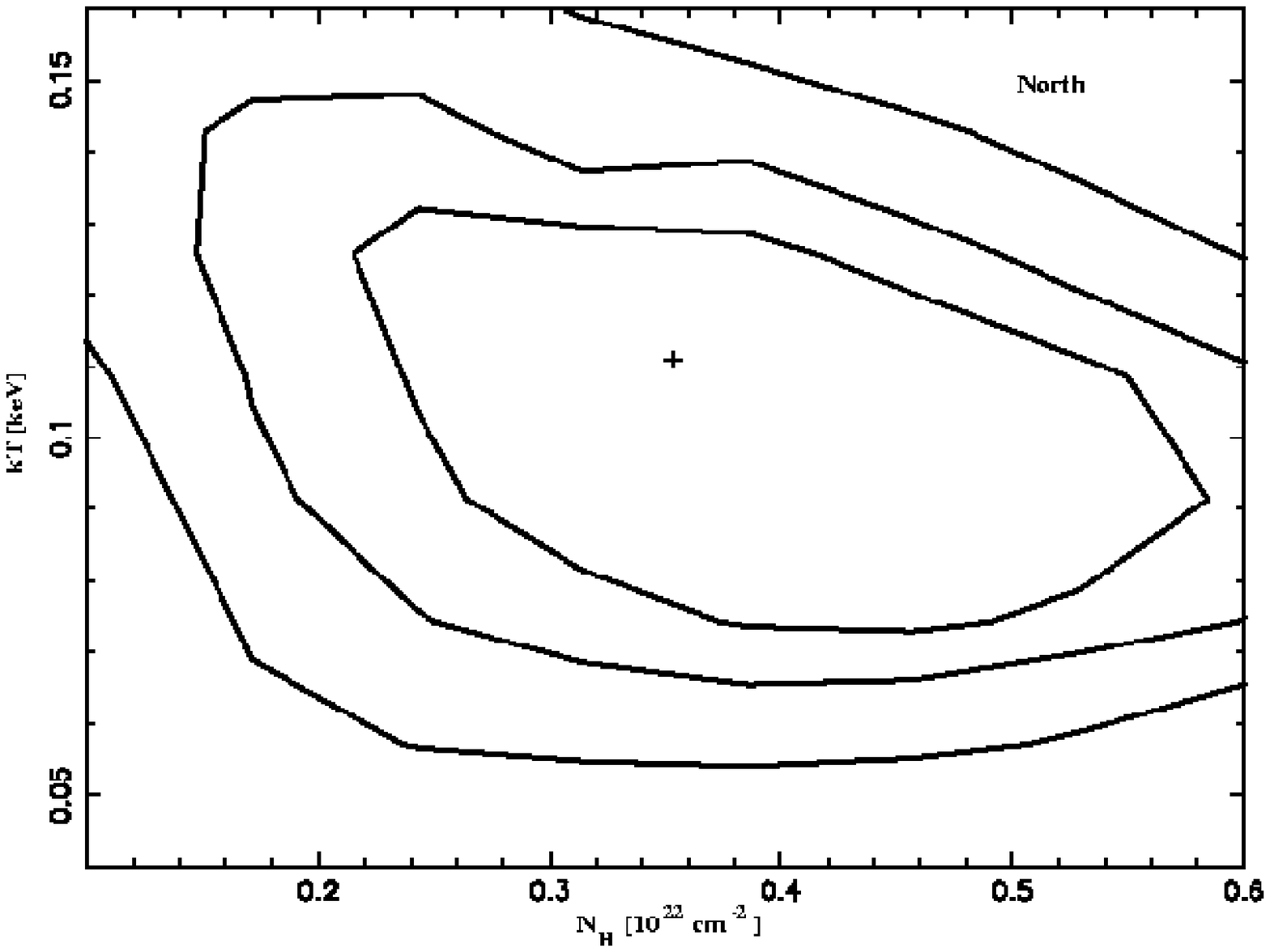}{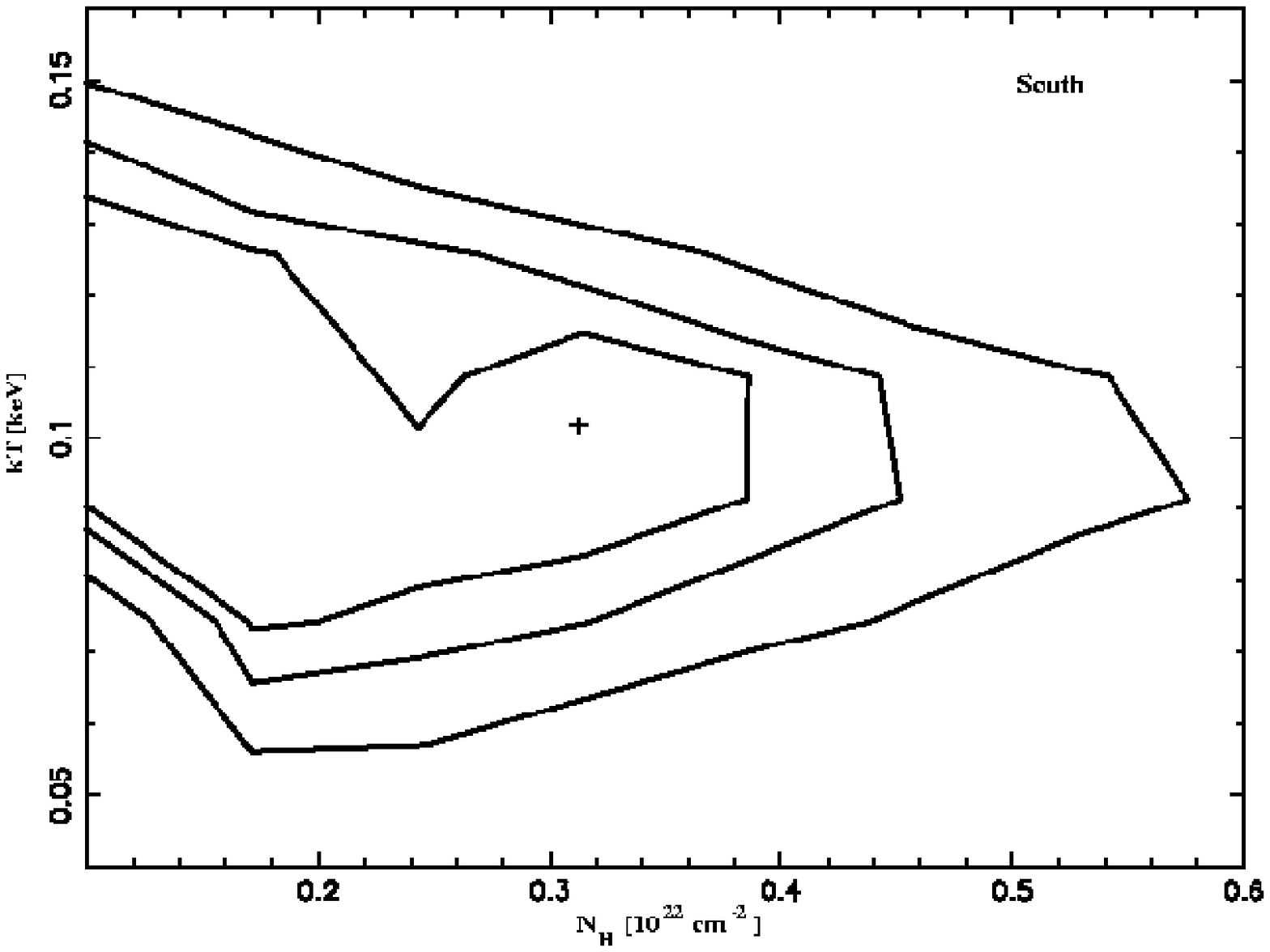}
\caption{\label{n-s-nh-t-chi2+f} $\chi^2$ contours for the best-fit 
results for spectra extracted from (a) the North Region and (b) the 
South Region. The free fit parameters are the temperature of the hot 
plasma component and the two normalization factors for the spectra.} 
\end{figure}

\begin{figure}  
\plotone{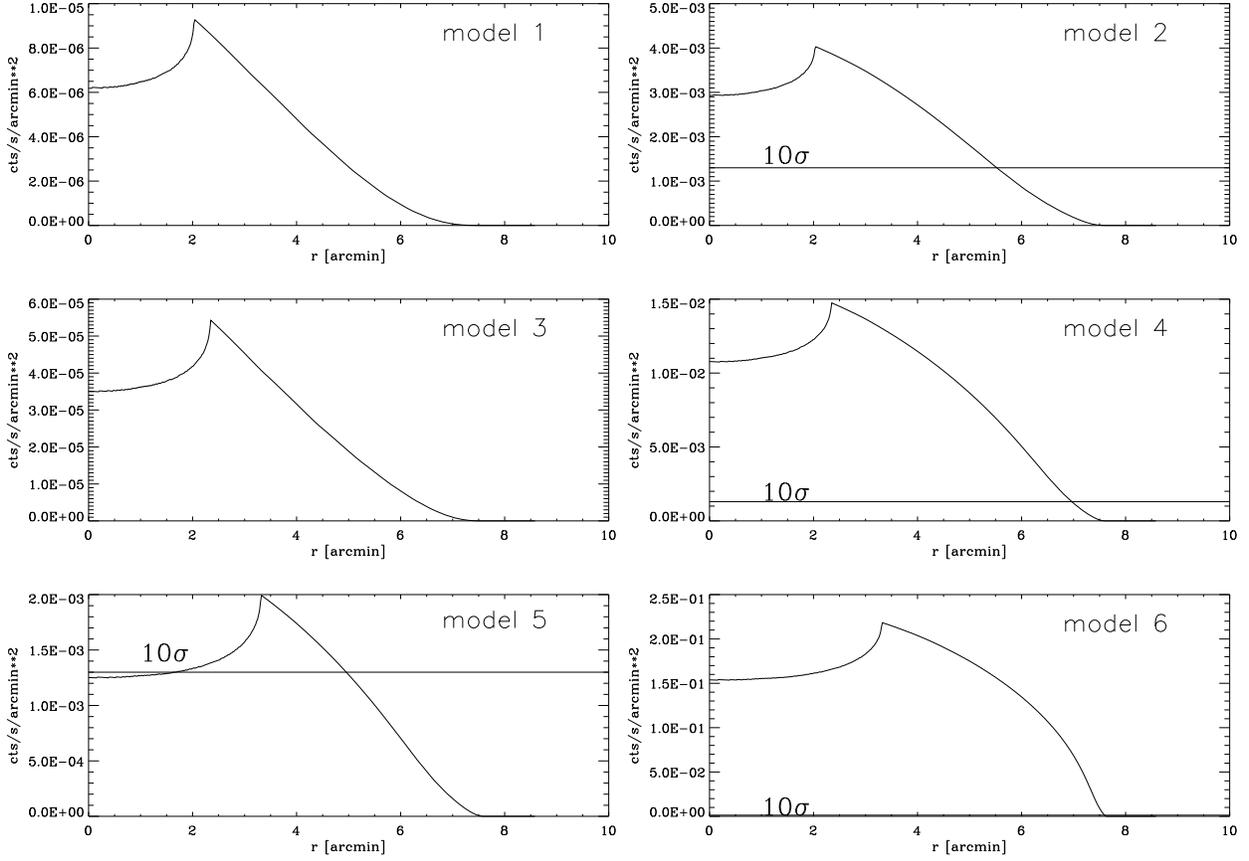}
\caption{\label{surf-model+f} Expected surface brightness profiles for
bubbles according to the analytical solution of \citet{GM95a}.
The surface brightness is expressed in {\it ASCA} 
counts~s$^{-1}$~arcmin$^{-2}$.
Although the entire {\it ASCA} energy band is used in this plot,
the effective upper cut off is imposed by the spectrum itself at
$\sim2$ keV.  
The bubble parameters of these six models are summarized in Table 4. 
Lines are drawn to show the 10$\sigma$ level in our {\it ASCA} SIS 
observations.  The radius of the outer shell is $\sim$7\farcm6.}
\end{figure}

\begin{figure} 
\plotone{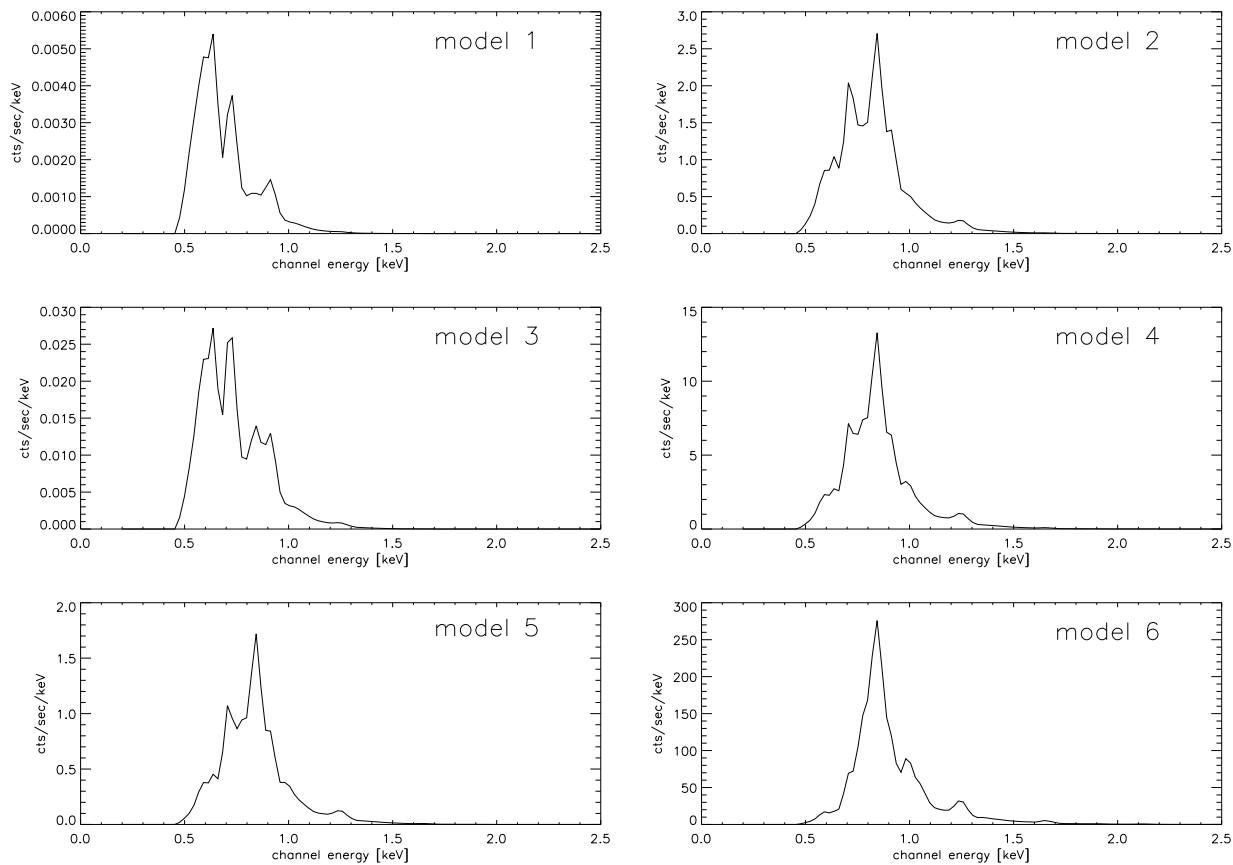}
\caption{\label{model-spec+f} Simulated {\it ASCA} SIS count rate
spectra for six bubble models calculated using the analytical solution 
of \citet{GM95a}.  The bubble parameters of these six models
are given in Table 4.  These simulated spectra can be compared
to the observed count rate spectra in Fig.~2.}
\end{figure}

\begin{figure}  
\plotone{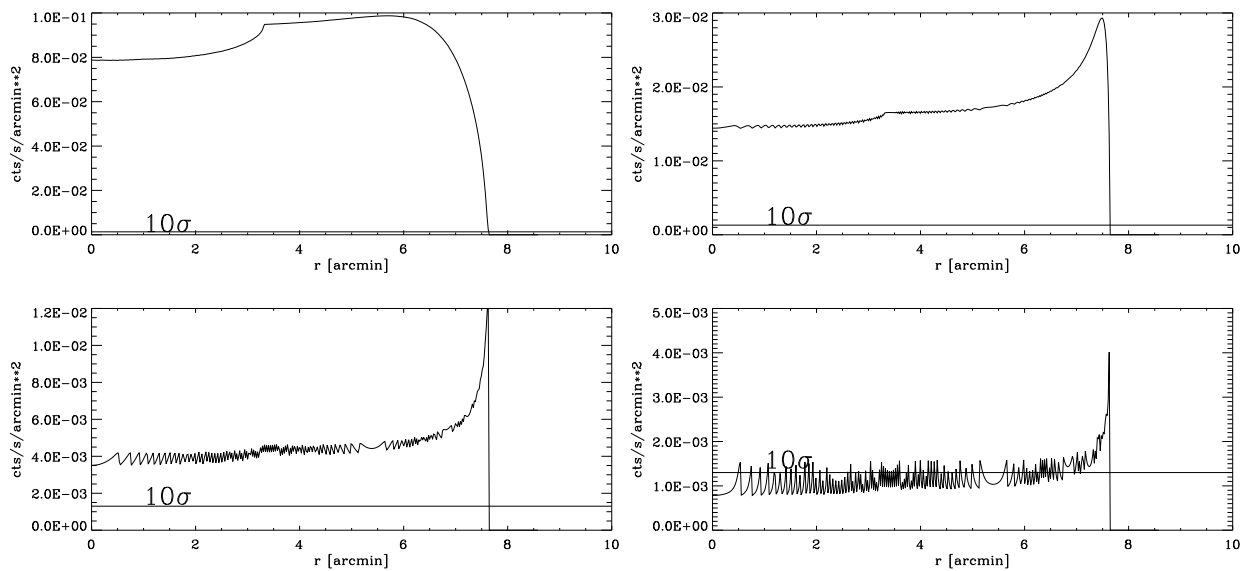}
\caption{\label{levap-model+f} Expected surface brightness for models
with decreased thermal conduction efficiency, i.e., increased
temperature scale. The temperature inside the bubble is increased by a
factor of 2 (upper left), 5 (upper right), 10 (lower left) and 20 
(lower right) compared to the initial model 6.  Lines are drawn to show
the 10$\sigma$ level in our {\it ASCA} SIS observations.}
\end{figure}

\clearpage
\begin{deluxetable}{cccccccc}
\tablewidth{0pt}
\tablecaption{Spectral Fits of the North and South Regions of 
NGC\,6888\tablenotemark{a}}
\tablehead{
Region                  &  $T_{\rm low}$   & $A_{\rm low}$\tablenotemark{b}    &
$T_{\rm high}$          & $A_{\rm high}$\tablenotemark{b}   &       &
Observed Flux           & Luminosity \\
Name                    &  (10$^6$ K)                     &   (cm$^{-5}$)          &
(10$^6$ K)              &  (cm$^{-5}$)             &    $\chi^2_{\rm red}$            &
(ergs~cm$^{-2}$~s$^{-1}$) &   (ergs~s$^{-1}$)
}
\startdata
North                   &   $1.3 \pm 0.1$    & $(7.6\pm0.4)$$\times$$10^{-3}$   &
$8.8 \pm 0.4$           &  $(2.1\pm0.3)$$\times$$10^{-4}$  &  1.8           &
4.8$\times$$10^{-13}$          & 1.4$\times$$10^{34}$ \\
South                   &  $1.3 \pm 0.1$     & $(1.1\pm0.4)$$\times$$10^{-2}$ &
$8.2 \pm 0.4$     &  $(1.7\pm0.3)$$\times$$10^{-4}$  &  1.7                   &
5.2$\times$$10^{-13}$          &  2.0$\times$$10^{34}$ \\
\enddata
\tablenotetext{a}{\citet{RS77} models were fitted to the {\it ASCA} SIS and 
{\it ROSAT} PSPC data simultaneously.}
\tablenotetext{b}{Normalization factor $A = 10^{-14}(\int{n_e^2 dV})/(4\pi\,D^2)$ 
in cgs units.}
\end{deluxetable}

\begin{deluxetable}{cccccccc}
\tablewidth{0pt}
\tablecaption{Spectral Fits of Three Concentric Elliptical Regions Regions of 
NGC\,6888\tablenotemark{a}}
\tablehead{
Region\tablenotemark{b} &  $T_{\rm low}$   & $A_{\rm low}$\tablenotemark{b}    &
$T_{\rm high}$          & $A_{\rm high}$\tablenotemark{c}   &       &
Observed Flux           & Luminosity \\
Name                    &  (10$^6$ K)                     &   (cm$^{-5}$)          &
(10$^6$ K)              &  (cm$^{-5}$)             &    $\chi^2_{\rm red}$            &
(ergs~cm$^{-2}$~s$^{-1}$) &   (ergs~s$^{-1}$)
}
\startdata
Ellipse 1               &   $1.3 \pm 0.1$    & $(1.5\pm0.4)$$\times$$10^{-2}$   &
$8.6 \pm 0.3$           &  $(4.2\pm0.4)$$\times$$10^{-4}$  &  1.56           &
9.9 $\times$$10^{-13}$          & 2.8$\times$$10^{34}$ \\
Ellipse 2               &  $1.1 \pm 0.1$     & $(2.7\pm1.1)$$\times$$10^{-2}$ &
$8.4 \pm 0.3$     &  $(2.3\pm0.3)$$\times$$10^{-4}$  &  1.28                   &
6.4$\times$$10^{-13}$          &  6.1$\times$$10^{34}$ \\
Ellipse 3               &  $1.3 \pm 0.2$     & $(4.1\pm2.9)$$\times$$10^{-3}$ &
$8.4 \pm 0.6$     &  $(8.2\pm1.7)$$\times$$10^{-5}$  &  1.23                   &
2.3$\times$$10^{-13}$          &  7.6$\times$$10^{33}$ \\
\enddata
\tablenotetext{a}{\citet{RS77} models were fitted to the {\it ASCA} SIS and 
{\it ROSAT} PSPC data simultaneously.}
\tablenotetext{b}{The three concentric elliptical regions are marked in Figure 1,
with Ellipse 3 being the innermost region.}
\tablenotetext{c}{Normalization factor $A = 10^{-14}(\int{n_e^2 dV})/(4\pi\,D^2)$ 
in cgs units.}
\end{deluxetable}

\clearpage

\begin{deluxetable}{llll}
\tablewidth{0pt}
\tablecaption{Physical Parameters of the WR Star and the Bubble Shell of NGC\,6888}
\tablehead{
Parameter\tablenotemark{a}    &  Value  & Unit   & Reference 
}
\startdata
$\dot{M}$        &  $6.3\times10^{-5}$ & $M_\odot$~yr$^{-1}$  & \citet{NCW98} \\
$v_{\infty}$      &  1600 & km~s$^{-1}$               & \citet{PBH90} \\
$L_{\rm w}$       &  5$\times$10$^{37}$&  ergs~s$^{-1}$& $L_{\rm w}=(1/2)\dot{M} v_{\infty}^2$ \\
$L_{\rm bol}$     & 1.6$\times$10$^{39}$ & ergs~s$^{-1}$ & \citet{vdH92} \\
$D$               & 1.8 & kpc                       & \citet{H78}   \\
$R_{\rm  shell}$  & 4   &  pc                       & measured from optical image\\ 
$v_{\rm exp}$     & 75$\pm$5 & km~s$^{-1}$          & \citet{TC82} \\
$M_{\rm ionized}$     & 5   & $M_{\odot}$               & \citet{Wen75} \\
$M_{\rm neutral}$ & 40  & $M_{\odot}$               & \citet{MM88} \\
\enddata
\tablenotetext{a}{$\dot{M}$ - mass loss rate; $v_{\infty}$ - terminal velocity 
of the fast stellar wind; $L_{\rm w}$ - mechanical luminosity of the stellar wind;
$L_{\rm bol}$ - bolometric luminosity of the star; $D$ - distance to NGC\,6888;
$R_{\rm  shell}$ - average shell radius; $v_{\rm exp}$ - shell expansion
velocity; $M_{\rm ionized}$ - ionized gas mass; $M_{\rm neutral}$ - neutral 
gas mass.}
\end{deluxetable}

\clearpage

\begin{deluxetable}{crrccrr}
\tablewidth{0pt}
\tablecaption{Six Models with Different Expansion Velocities and Shell Masses}
\tablehead{
Model & $v_{\rm exp}$ & $M_{\rm shell}$ & $L_{\rm w}$ &  $L_{\rm x}$ &  Count Rate \\ 
Number&  (km~s$^{-1}$) & ($M_\odot$)   & (erg~s$^{-1}$) &  (erg~s$^{-1}$) & 
(cnt~s$^{-1}$)
}
\startdata
1       &75     &4      & 4.1$\times$$10^{35}$  & 1.2$\times$$10^{33}$ & 5.3$\times$$10^{-4}$ \\
2       &75     &40     & 4.1$\times$$10^{36}$  & 8.7$\times$$10^{34}$ & $0.3$ \\
3       &100    &4      & 9.7$\times$$10^{35}$  & 3.6$\times$$10^{33}$ & 3.4$\times$$10^{-3}$ \\
4       &100    &40     & 9.7$\times$$10^{36}$  & 2.1$\times$$10^{35}$ & $1.3$ \\
5       &200    &4      & 7.7$\times$$10^{36}$  & 3.6$\times$$10^{34}$ & $0.17$        \\
6       &200    &40     & 7.7$\times$$10^{37}$  & 1.6$\times$$10^{36}$ & $25.2$        \\
\enddata
\end{deluxetable}

\end{document}